\documentclass{appolb}

\usepackage{epsfig}

\newcommand{\beq}{\begin{equation}}
\newcommand{\eeq}{\end{equation}}
\def\frac#1#2{{#1 \over #2}}

\def\half{\ifinner {\scriptstyle {1 \over 2}}
   \else {1 \over 2} \fi}

\def\simge{\mathrel{%
   \rlap{\raise 0.511ex \hbox{$>$}}{\lower 0.511ex \hbox{$\sim$}}}}
\def\simle{\mathrel{
   \rlap{\raise 0.511ex \hbox{$<$}}{\lower 0.511ex \hbox{$\sim$}}}}

\def\del{\partial}

\begin{document}

\title{Saturation: Colour Glass Condensate and colour sources
\thanks{Presented at XXXIV International Symposium on 
Multiparticle Dynamics\\ July 26-August 1, 2004, Sonoma County, California, USA}
}
\author{Elena G. Ferreiro
\address{Departamento de F\'{\i}sica de Part\'{\i}culas,
%Universidad de
%Santiago de Compostela,
15706 Santiago de Compostela, Spain \\
E-mail: elena@fpaxp1.usc.es}
}
\maketitle

\begin{abstract}
%I review the phenomenon of parton saturation at small $x$ in the
%framework of the Colour Glass Condensate, and I discuss some of its applications
%at RHIC energies.
In the recent years we have seen a lot of activity around systems and experiments,
like DIS at HERA or the heavy-ion experiments at RHIC, involving a large number of partons
due to the high enery and/or the high number of participants of those experiments.
The main problem in this regime is that of the high parton densities.
In fact, in most of the models of multiparticle production,
two contributions to the multiplicity are considered: one proportional to the number of
participant nucleons, $N_{part}$, and a second one proportional to the number of collisions,
$N_{part}^{4/3}$. In order to get the right multiplicities at RHIC it is necessary to lower
the second contribution. A possible mechanism for this is the saturation.
Here, I am going to review the saturation of parton densities
in the initial state, in two different frameworks: the Colour Glass Condensate %(CGC)
and the string clustering.
\end{abstract}

%\section{Introduction: Parton saturation at small $x$}
%In the recent years we have seen a lot of activity around systems and experiments,
%like DIS at HERA or the heavy-ion experiments at RHIC involving a large number of partons
%due to the high enery and/or the high number of participants of those experiments.
%The main problem in this regime is that of the high parton densities.
%In fact, in most of the models of multiparticle production 
%two contributions to the multiplicity are considered: one proportional to the number of 
%participant nucleons, $N_{part}$, and a second one proportional to the number of collisions,
%$N_{part}^{4/3}$. In order to get the right multiplicities at RHIC it is necessary to lower
%this second contribution. Here, I am going to review the mechanism of saturation
%in the initial state, in two different frameworks: the Colour Glass Condensate (CGC)
%and the string clustering. 

\section{THE COLOUR GLASS CONDENSATE}

\subsection{Parton saturation at small $x$ and the saturation momentum}

It has been much activity in the last years trying 
to understand the physics of nuclear
and hadronic processes in the regime of very small Bjorken's $x$ (very high
energy). The main problem in this regime is that of the high parton densities.
At high energy, the QCD cross sections are controlled by small longitudinal
momentum gluons in the hadron wavefunction, whose density grows rapidly with
increasing energy or decreasing $x$, due to the enhancement of radiative
process. If one applies perturbation theory to this regime, one finds
that, by resumming dominant radiative corrections at high energy, the BFKL
equation leads to a gluon density that grows like a power of $s$ and in
consequence
to a cross section that violates the Froissart bound.
Nevertheless,
 the use of perturbation theory to high-energy problems is not obvious. In
fact, the BFKL and DGLAP equations are linear equations that neglet the
interaction among the gluons. With increasing energy, recombination effects
--that are non-linear--
favored by the high density of gluons should become more important and lead
to an eventual {\it saturation} of parton densities. 
%At these energies, the gluon density in the hadron wavefunction would be so
%high that the perturbation theory would break down, even for small coupling
%constant, due to strong non-linear effects. 

These effects become important
when the interaction probability for the gluons becomes of order one.
Taking $\frac{\alpha_s N_c}{Q^2}$ as the transverse size of the gluon and 
$\frac{x G(x,Q^2)}{\pi R^2}$ as the density of gluons, the interaction
probability is expressed by
\beq
\frac{\alpha_s N_c}{Q^2}\,\,\times\,\,
\frac{x G(x,Q^2)}{\pi R^2}\ .
\label{ec1}
\eeq
Equivalently, for a given energy, saturation 
occurs for those gluons having a sufficiently large transverse size $r_\perp^2
\sim 1/Q^2$, larger than some critical value $1/Q_s(x,A)$. So the phenomenon
of saturation introduces a characteristic momentum scale,
the {\it saturation momentum} $Q_s(x,A)$, which is a measure of the
density of the saturated gluons, and grows rapidly with $1/x$ and
$A$ (the atomic number). 
The probability of interaction 
--that can be understood as "overlapping" of the gluons in the transverse space--
becomes of order one for those gluons with
momenta $Q^2 \simle Q_s(x,A)$ where 
\beq
Q^2_s(x,A)=\alpha_s N_c \ \frac{x G(x,Q^2_s)}{\pi R^2} \equiv
\frac{({\rm colour\,\, charge})^2}{{\rm area}}\ .
\label{ec2}
\eeq

%For $Q^2\gg Q^2_s(x,A)$,  evolution equations (like BFKL or DGLAP) apply.
%In particular, one can estimate the saturation scale by
%inserting the BFKL approximation into
%eq.~(\ref{ec2}). This gives:
%% \cite{AM2,SCALING} :
%\beq
%\label{ec3}
%Q_s^2(\tau)\,=\,\Lambda^2
%{\rm e}^{c\bar\alpha_s \tau}\,,\qquad c\,=\,%\frac{1}{2}
%\big[-{\beta}+\sqrt{\beta(\beta+8\omega)}\,\big]/2\,=\,4.84...\ ,
%\eeq where
%$\tau\equiv \ln(1/x)$.

For $Q^2\simle Q^2_s(x,A)$, the non-linear effects
are essential, since they
 are expected to soften the growth of the gluon distribution
with $\tau\equiv \ln(1/x)$.
For a nucleus, $x G_A(x,Q^2_s)\propto A$ and
 $\pi R^2_A\propto A^{2/3}$, so eq.~(\ref{ec2})
predicts
$Q^2_s\propto A^{1/3}$. One can estimate the saturation scale by
inserting the BFKL approximation into
eq.~(\ref{ec2}). This gives
%To summarize 
(with $\delta\approx 1/3$ and $\lambda\approx c\bar\alpha_s$ in a
first
approximation):
\beq
\label{ec4}
Q^2_s(x,A)\,\,\sim\,\,A^{\delta}\, x^{-\lambda}\,,
\qquad c\,=\,\big[-{\beta}+\sqrt{\beta(\beta+8\omega)}\,\big]/2\,=\,4.84...\ ,
\eeq
which indicates that an efficient way to create a high-density
environment is to combine large nucleai with moderately small values of $x$,
as it is done at RHIC.

This equation also shows that for sufficiently large energy or $x$ small
enough, $Q^2_s(x,A) \gg \Lambda^2_{QCD}$ and $\alpha_s(Q_s) \ll 1$, which
characterizes 
the regime of weak coupling QCD. But although the coupling
is small, the effects of the interactions are amplified by the large gluon
density: at saturation, $ G_A(x,Q^2_s) \sim 1/\alpha_s(Q_s) \gg 1$, so 
the gluon modes
have large occupation numbers, of  order $1/\alpha_s$ (corresponding
to strong classical fields $A\sim 1/g$),
which suggests the use of semi-classical methods.

\subsection{The effective theory for the CGC: the Renormalization Group Equation.}
%Saturation}

One can write a classical effective theory based on this general idea: the
fast partons -valence quarks with large longitudinal momentum- are considered
as a {\it classical source} that emits soft gluons
%\footnote{The soft gluons 
%-classical colour fields- see the
%fast partons -random colour source- as an effective colour charge which is
%static, since the gluons have large energies and shorter lifetimes.}
 -with smaller longitudinal
momenta- which are treated as {\it classical colour fields} ${A[\rho]}$.
The fast partons move nearly at the speed of light in the positive $x^+$
direction, and generate a colour current
$J^\mu=\delta^{\mu +}\!\rho$. By Lorentz contraction, the support
of the charge density $\rho$ is concentrated near the light-cone longitudinal
coordinate $x^-=0$. By Lorentz time
dilation, $\rho$ is independent of the
light-cone time 
$x^+$.
%\footnote{I use
%light-cone vector notations: $p^-$= light-cone energy, $p^+$= light-cone
%longitudinal momentum, $x^+$= light-cone time, $x^-$= light-cone longitudinal
%coordinate}.

The Yang Mills equation describing the soft gluon dynamics reads
\beq
D_\nu F^{\nu \mu}\,\,=\,\,\delta^{\mu +}\rho(x^-,{\bf x})\ .
\label{ec5}
\eeq
Physical quantities, as the unintegrated gluon distribution, are obtained as
an average over $\rho$:
\beq
< A^i(X) \, A^i(Y) >_x\,=\int D[\rho]\,\,W_x[\rho]\,\,
{A}^i[\rho](X)\,{A}^i[\rho](Y)\ ,
\label{ec6}
\eeq
where $A^i(X)$ corresponds to the classical solution for a given $\rho$,
and $W_x[\rho]$ is a gauge-invariant weight function for $\rho$.
What we are doing is a kind of
Born-Oppenhaimer approximation:
first, we study the dynamics of the classical
fields (Weizsacher-William fields)
for a given configuration $\rho$ of the colour charges,
and second,
we average over all possible configurations.

For the classical solution we find:
\beq
F^{+i}(x^-,x_\perp)= \delta(x^-) \frac{i}{g} V(x_\perp) (\partial^i
V(x_\perp)^\dagger)=\partial^+ A^i\ , 
\label{ec7a}
\eeq
where $V(x_\perp)$ is the Wilson line
\beq
V^\dagger(x_\perp)\equiv{\rm P} \!\exp
 \Big\{
ig\! \int\! dx^- \!A^+(x^-,{\bf x})
 \Big\}
\label{ec7}
\eeq
and $A^+[\rho]$ is the solution of the equation of motion (\ref{ec5}) in the
covariante gauge: $- \nabla^2_\perp A^+=\rho$.

%Concerning the weight function, $W_\tau[\rho]$ is obtained by integrating out the
%fast
%partons, so it depends upon the rapidity scale $\tau={\rm ln}(1/x)$ at which
%one considers the effective theory.
% With increasing energy (or decreasing x) new quantum modes become relatively
%"fast" and must be included in the colour source seen by the gluons. Thus, the
%classical
%description of the small-$x$ gluons is to be seen as an effective theory valid
%at a given
%value of $x$, and whose "action" is evoluting with $x$.
%This can be done via a one-loop background field calculation, and
%leads to a renormalization group equation (REG) for $W_\tau[\rho]$
%which shows how the correlations of $\rho$ change with increasing $\tau$.
%%\cite{PI,JKLW97}.
The weight function $W_\tau[\rho]$ is obtained by integrating out the
fast
partons, so it depends upon the rapidity scale $\tau={\rm ln}(1/x)$ at which
one considers the effective theory.
This can be taken into account via a one-loop background field calculation, and
leads to a renormalization group equation (REG) for $W_\tau[\rho]$
which shows how the correlations of $\rho$ change with increasing $\tau$.
Schematically:
\beq
{\del W_\tau[\rho] \over {\del \tau}}\,=\,
 {1 \over 2} \int_{x_\perp,y_\perp} \,{\delta \over {\delta
\rho_x^a}}\,\,\chi_{xy}^{ab}[\rho]\,\,{\delta \over {\delta\rho_y^b}}\,\,
W_\tau[\rho]\ .
\label{ec8}
\eeq
This is a functional diffusion equation, where the kernel
$\chi[\rho]$ plays the role of the
diffusion coefficient in the functional space spanned by
$\rho(x^-,x_\perp)$. 
%This kernel is positive definite and non-linear in $\rho$
%to all orders.
%This equation encompasses previous evolution equations \cite{BKW}.
%developed by Balitsky,
%Korchegov and Weigert.
This kernel is positive definite and non-linear in $\rho$
to all orders. It depends upon $\rho$ via the Wilson line (\ref{ec7}).

As it has been said above, the physical quantities, as the gluon  density, are
obtained as an
average over $\rho$:
\beq
n(x, k_\perp)\,\equiv\,\frac{1}{\pi R^2}\,
\frac{{\rm d}N}{{\rm d}{\tau}\,{\rm d}^2 k_\perp}\,\propto\,
\langle F^{+i}(k_\perp) F^{+i}(-k_\perp)\rangle_{x}\ .
\label{ec9}
\eeq
In order to calculate this average, we need to solve the
REG (\ref{ec8}).
Approximate solutions to this equation can be obtained in two limiting cases:

$\bullet$ At low energy, or large transverse momenta $k_\perp^2 \gg Q_s^2(x)$,
we are in a dilute regime where fields and sources are weak, and
the Wilson lines can be expanded to lowest order in $A^+$,
$V^\dagger(x_{\perp})\approx 1 + ig \int dx^- A^+ (x^-,x_{\perp}) $.
In this case, the REG equation reduces to the BFKL equation, and the gluon
density of eq. (\ref{ec9}) grows both with $1/k_\perp^2$
and $1/x$ (Bremsstrahlung):
%\beq
$n(x,k_\perp)\,\sim\,\frac{1}{k_\perp^2}\,\frac{1}{x^{\omega\alpha_s}}$.%\ .
%\label{ec10}
%\eeq

$\bullet$ At high energies, or low momenta $k_\perp^2 \simle Q_s^2(x)$,
the colour fields are strong, $A^+ \sim 1/g$, so the
Wilson lines rapidly oscillate and
average away to zero: $V\approx V^\dagger \approx 0$. Then the
kernel $\chi$ becomes independent of $\rho$, and we obtained a gluon density
that increases linearly with the evolution "time" $\tau=ln(1/x)$:
%\beq
$n(x,k_\perp)\,\sim\,{1\over \alpha_s}\,
\ln{Q_s^2(x)\over k_\perp^2}\, \propto \,\,\ln{1\over x}$. %\ .
%\label{ec11}
%\eeq
That is, we find {\it saturation} for the gluon density, that grows
logaritmically with the energy since $\tau \sim \ln s$: unitarity is
restored.
%\begin{figure}[t]
%%\centering\leavevmode
%\centerline{\epsfxsize=2.5in\epsfysize=2in\epsffile{SAT3.eps}
%\epsfxsize=2.5in\epsfysize=2in\epsffile{SAT4.eps}}
%\caption{(a) Gluon phase-space density as a function of $k_\perp$;
%note the change in behaviour below $Q_s$.
% (b) The same as (a), but for two values of $x$.}
%\end{figure}

We call the high density gluonic matter at small-$x$ described by this
effective theory a {\it Colour Glass Condensate} (CGC) \cite{PI}:
{\it Colour } since gluons carry colour under $SU(N_c)$;
{\it Glass } since we have a random distribution of time-independent
colour charges
which is averaged over in the calculation
of physical observables, in order to have a gauge independent formulation;
and {\it  Condensate } because at saturation the gluon density is of
order $1/\alpha_s$, typical of condensates, so we have a system of saturate
gluons that is a Bose condensate.

\subsection{Phenomenology at RHIC}

The particle production in RHIC collisions has been analyzed from the
perspective of the CGC,
considering it as a pertinent description
of
the initial conditions. Taking into account that the multiplicity is
proportional to the number of gluons -parton-hadron duality-, the centrality
dependence of multiparticle production has been related to the density
of gluons \cite{KNL}, that at saturation (see eq. (\ref{ec2})):
\beq
\frac{{\rm d} N}{{\rm d}y}\,\sim\, 
x G(x, Q_s^2)\,=\,\frac{{\pi R^2_A}\, Q_s^2(x,A)}{\alpha_s( Q_s^2)}
\label{ec13}
\eeq
%where $c$ refers to the fractions of gluons liberated in the collision and $R$
%is a multiplication factor that takes into account
% the inelasticity in the evolution of the parton system towards equilibrium. 
%
%At saturation (see eq. (\ref{ec2})),
%\beq
%x G(x, Q_s^2)\,=\,\frac{{\pi R^2_A}\, Q_s^2(x,A)}{\alpha_s( Q_s^2)}
%\label{ec13}
%\eeq
where
$\pi R^2_A \,\propto\,N_{part}^{2/3}$ corresponds to the nuclear overlap area,
and $Q_s$ is the saturation momentum for the considered centrality, 
$Q_s^2(x,A)\,\propto\,N_{part}^{1/3}$.
To compute the centrality dependence, it is necessary to know the evolution of
the gluon structure function, wich is governed by the DGLAP equation. Taking
%\beq
$1/\alpha_s( Q_s^2)\,\approx\, \ln( Q_s^2/\Lambda^2_{QCD})\ 
\,\sim\,\ln N_{part}$, %\ ,
%\label{ec14}
%\eeq
one finally finds that the multiplicity per participant behaves as 
$\ln N_{part}$.
%\beq
%\frac{2}{N_{part}}\,
%\frac{{\rm d} N}{{\rm d}{\rm y}}\,\sim\, %c\, R\,
%\frac{1}{\alpha_s( Q_s^2)}\,\sim\, %c\, R\, 
%\ln N_{part}\ .
%\label{ec15}
%\eeq

%Concerning the energy dependence \cite{KL}, it comes via the saturation scale
%$Q_s^2(x) = Q_0^2 \ (x_0 / x)^{\lambda}$, where $\lambda$ is obtained from the
%saturation fits to $F_2$ at HERA, $\lambda \simeq 0.25 \div 0.3$.
%Then, we can relate the multiplicities at 200 GeV with the ones at 130 GeV
%by:
%%%\beq
%%\begin{eqnarray}
%%{dN \over d \eta} (\sqrt{s} =
%% 200 \rm{GeV}) &\!\!\simeq\!\!& \left({200 \over 130}\right)^{\lambda}
%%{dN \over d \eta} (\sqrt{s}=130 \rm{GeV}) &=&  (1.10 \div 1.14) 
%%{dN \over d \eta} (\sqrt{s}
%%= 130 \rm{GeV})\ ,
%%\label{ec16}
%%\end{eqnarray}
%%%\eeq
%%\beq
%\begin{eqnarray}
%{dN \over d \eta} (\sqrt{s} =
% 200 \rm{GeV}) \simeq\ \left({200 \over 130}\right)^{\lambda}
%{dN \over d \eta} (\sqrt{s}=130 \rm{GeV})\\ \nonumber =  (1.10 \div 1.14)
%{dN \over d \eta} (\sqrt{s}
%= 130 \rm{GeV})\ ,
%\label{ec16}
%\end{eqnarray}
%%\eeq
%that agrees with the Phobos results: $R(200/130) = 1.14\pm 0.06$
%
%%%%%%%%%%%%
Besides, it has been obtained from saturation a proportionality between the mean 
transverse momentum and the multiplicities, 
\beq
<p_T>^2 \sim \frac{1}{\pi R_A^2 }
\frac{{\rm d} N}{{\rm d}y}\ .
%\frac{dN}{dy}\ .
\label{ec17}
\eeq 
This observation indicates that the $p_T$ broadening seen in elementary
and heavy ion collisions results from the same physics, the intrinsic generated
$p_T$ broadening in the partonic phase \cite{SB01}.
%The idea of geometric scaling at saturation described above has also been
%applied to the phenomenology
%of heavy ion collisions at RHIC. 
%In Ref. \cite{SB01}, it has been argued
%that the RHIC data for the
%transverse momentum distributions of the produced hadrons show
%geometric scaling in their dependences upon $m_T$ and centrality:
%they are universal functions of $m_T/\Lambda_s(b)$. % with
%%$b$ the impact parameter.

%In conclusion, we can say that the saturation picture based on the CGC can
%explain, at least partially, some of the RHIC results, as 
%the weak dependence of the multiplicity per
%participant
%on the number of
%participants or the high $p_T$
%suppression.

\section{STRING CLUSTERING}

In the clustering approach, the colour strings created in the
nuclear collisions are considered 
as effective sources with a fixed transverse area, $r_\perp \approx 0.2$ fm.
Notice that this radius coincides with the estimate saturation momentum
of the CGC at RHIC. %($Q_s^2 \sim 1....2$ GeV$^2$).
If the string density is high enough, the strings {\it overlap},
forming clusters \cite{ABFP},
very much like disks in continuum two-dimensional percolation
theory. In order to calculate the
physical observables, as the particle multiplicity
or the mean transverse momentum, we need to study the dynamics of those
clusters.

We assume that a cluster of $n$ strings
behaves as a single string with a higher colour field $\vec Q_n$,
corresponding to the vectorial sum of the colour charges of each individual
$\vec Q_1$ string. The resulting colour field covers the area $S_n$ of
the cluster. As $\vec Q_n^2=( \sum_1^n \vec Q_1)^2$,
and the individual
string colours may be oriented in an arbitrary manner, %respective to one
%another, 
the average $\vec Q_{1i}\vec Q_{1j}$ is zero, so
$\vec Q_n^2=n \vec Q_1^2$.
$\vec Q_n$ depends also on the area $S_1$ of each individual string
that comes into the cluster, as well as on
the total area of the cluster $S_n$, $Q_{n}= \sqrt{ n S_n \over S_1} Q_{1}$.
%\begin{equation}Q_{n}= \sqrt{ n S_n \over S_1} Q_{1}\ .
%\label{ec11n}\end{equation}
% CHANGE 6
We take $S_1$
constant and equal to a disc of radius $r_\perp \simeq 0.2$ fm.
$S_n$ corresponds to the total area occupied by
$n$ discs\footnote{
Notice that if the strings don't overlap, $S_n=n S_1$
and $Q_{n}=n Q_{1}$,
so
the strings act independently. On the contrary, if they
fully overlap,
$S_n=S_1$ and $Q_{n}=\sqrt {n}Q_{1}$.
}.
Knowing the colour charge $Q_n$, 
one can compute the multiplicity $\mu_n$ and
the mean transverse momentum $<p_T^2>_n$ of the particles produced by a cluster
of $n$ strings. One finds \cite{BMP}
\beq
\mu_n=\sqrt{\frac{n S_n}{S_1}} \mu_1\ ,\ \ \ \ \ \ <p_T^2>_n=\sqrt{\frac{n
S_1}{S_n}} <p_T^2>_1
\label{ec12n}
\eeq
where $\mu_1$ and $<p_T^2>_1$ are the mean multiplicity and mean $p_T^2$
of particles produced by a simple string.
In the saturation limit, i. e. all the strings overlap into a single cluster
that occupies the whole interaction area, one gets the following scaling law
that relates the mean transverse momentum and the multiplicity per unit
rapidity and unit transverse area:
\beq
<p_T^2>_{AA}=\frac{S_1}{S_{AA}} \frac{<p_T^2>_1}{\mu_1} \mu_{AA}\ .
\label{ec13n}
\eeq
% CHANGE
This scaling relation is similar to the one obtained in the framework of the
CGC
when the initial gluon density saturates.

Moreover, in the limit of high density $\eta= N_{s} S_1/S_{AA}$,
one obtains
\beq
<\frac{n S_1}{S_n}>= \frac{\eta}{1-\exp{(-\eta)}} \equiv \frac{1}{F(\eta)^2}
\label{ec14n}
\eeq
and
the equations (\ref{ec12n}) transform into the analytic
ones \cite{DFPU}
\beq
\mu= N_s F(\eta) \mu_1\ ,\ \ \ \ \ \ <p_T^2>=\frac{1}{F(\eta)}
<p_T^2>_1
\eeq
where $\mu$ and $<p_T^2>$ are the total multiplicity and mean momentum, and
$N_s$ is the total number of strings created in the collision.

In order to study the transverse momentum distribution,
one needs the distribution $f(x,m_T)$ for each string or cluster, and the
cluster size distribution $W(x)$.
For $f(x,m_T)$ we assume
the Schwinger formula, $f(x,m_T)=\exp(-m_T^2 x F(\eta))$, used also
for the fragmentation of
a Lund string. In this formula $x$ is related
to the string tension, or equivalently to the mean transverse
size
%mass
of the
string. The weight function $W(x)$ obeys a gamma distribution \cite{DFPU}.
%depends on the area of the string.
Assuming that a cluster behaves similarly to a single string
but with different string tension, that depends on the number of strings that
come into the cluster,
%and taking into account that the cluster tension and the cluster size behaves
%in a similar way, since both depend equivalently on the number of participant
%strings
%that come into each cluster,
we can write for the total $p_T$ distribution
\beq
f(m_T)=\int W(x)\ f(x,m_T)\,,\qquad W(x)=\frac{\gamma}{\Gamma(k)} (\gamma x)^{k-1}\ \exp{(-\gamma
x)}\ .
\label{ec2n}
\eeq
%where the weight function $W(x)$ %, which reflects the cluster size distribution,
%obeys a gamma distribution.
%\beq
%W(x)=\frac{\gamma}{\Gamma(k)} (\gamma x)^{k-1}\ \exp{(-\gamma x)}\ .
%\label{ec3n}
%\eeq
Performing the integral in eq. (\ref{ec2n}) we obtain
\beq
%\frac{A}{(1 +{F(\eta)\ p_T^2 \over \gamma})^k}\
\frac{{\rm d} N}{{\rm d} p_T^2 {\rm d} y}=\frac{{\rm d} N}{{\rm d}y} 
\frac{k-1}{\gamma} F(\eta)
\frac{1}{(1 +\frac{F(\eta)\ p_T^2}{\gamma})^k}\ ,
\label{ec8n}
\eeq
where $k$ can be obtained from the fluctuations in multiplicy.
%\beq
%k= \frac{2 \phi^2(\eta)}{\phi^2(\eta) -1}
%\label{ec9n}
%\eeq
%with $\phi(\eta)= \sqrt{F(\eta)} + \sqrt{\eta}\ F(\eta)$.
%%Our results from this formula are compared to PHENIX data on $\pi^0$ production for
%%%central p-p,
%%central Au-Au and peripheral Au-Au collisions at $\sqrt{s}=200$ GeV in Fig. \ref{fig2new}.

\section{SOME SIMILARITIES: The results from string clustering and from
the CGC}

To finish, let us remember some of the similarities of the explained
approaches:

%\vspace*{-.1cm}
%\begin{itemize}
%\item {
\vspace*{0.2cm}
$\bullet$ In the clustering approach, when taking the saturation limit
--all the
strings overlap into a single cluster that occupies the whole nuclear overlap
area--, 
one finds that the particle multiplicity of a central collision,
$\mu_{AA}$, behaves as $\mu_{AA}=\mu_n=\sqrt{\frac{n S_n}{S_1}} \mu_1=
\sqrt{\frac{N_s S_{AA}}{S_1}} \mu_1$.
%\beq
% \mu_{AA}=\mu_n=\sqrt{\frac{n S_n}{S_1}} \mu_1=
%\sqrt{\frac{N_s S_{AA}}{S_1}} \mu_1\ .
%\label{ec17n}
%\eeq     
Taking into account that the number of strings produced in the nuclear
collision, $N_s$, is proportional to the number of inelastic nucleon-nucleon
collisions, $N_{coll} \sim A^{4/3}$, and $ S_{AA}$ corresponds to the nuclear
overlap area, $ S_{AA} \sim A^{2/3}$, we obtain a multiplicity that scales with
the number of participants $A$. This coincides with the multiplicity obtained
in first approximation -- without evolution -- in the framework of the
CGC (eq.~(\ref{ec13})).
%\beq
%\frac{d N}{d y}\,\sim\,  x G(x, Q_s^2)\,\sim\,
%\frac{{\pi R^2_A}\, Q_s^2(x,A)}{\alpha_s( Q_s^2)}
%\label{ec100}
%\eeq
%where
%$\pi R^2_A \,\propto\,A^{2/3}$ corresponds to the nuclear overlap area,
%and $Q_s$ is the saturation momentum for the considered centrality,
%$Q_s^2(x,A)\,\propto\,A^{1/3}$.
%}
%\item {

\vspace*{0.2cm}
$\bullet$
In the CGC, there is a relation between the mean transverse momentum and
the multiplicity, which is developed in eq.~(\ref{ec17}).
This relation shows that at saturation the mean transverse momentum should
scale with the multiplicity per unit rapidity and unit transverse area.
This coincides with
the proportionality relation obtained in the clustering model
(eq.~(\ref{ec13n})).%}

\vspace*{0.2cm}
%\item {
$\bullet$
In both approaches, the initial state interactions --gluon saturation
in the CGC or clustering of strings--
%in the percolation approach -
produce a suppression
%of the $p_T$ distributions.
%In the case of the CGC this suppression modifies the hard perturbative QCD
%$p_T$ distributions.
%In the percolation approach, the clustering of soft strings reproduces the
%shape of the $p_T$ distributions, whose slope is larger at higher density.
%%is produced up to the scale of
%%saturation ($Q_s(x) \simeq 2$ GeV$^2$ at RHIC) and it is preserved by quantum
%%evolution up to 6 GeV$^2$.
%In both approaches there is a suppression
of
high $p_T$ and multiplicities.
On the contrary, in the framework of the jet quenching phenomena,
the energy loss of the jet with a hot and dense medium produces additional
soft gluons that would fragment into hadrons increasing the multiplicities,
unless strong shadowing occurs in the gluon structure functions.
%}
%\end{itemize}

\end{document}